# Intrinsic physical properties of flexible van der Waals semiconductor InSe

Jacob Svane,[a] Khuong Kim Huynh,[a*] Yong P. Chen,[b] and Bo Brummerstedt Iversen[a*]

[a]Center for Sustainable Energy Materials, Department of Chemistry, Aarhus University, 8000 Aarhus C, Denmark

[b]Department of Physics and Astronomy, Aarhus University, 8000 Aarhus C, Denmark

E-mail: hkkhuong@chem.au.dk, bo@chem.au.dk

**ABSTRACT**

**InSe is a van der Waals semiconductor in which mechanical flexibility, high electronic mobility, and non-trivial electronic structures converge, making it an attractive platform for both intriguing fundamental studies and promising device developments. However, the nucleation and growth of phase-pure, intrinsic InSe crystals require stringent thermodynamical conditions, and have therefore remained elusive. Since InSe melts incongruently, the widely used synthesis methods based on cooling of a 1:1 In–Se mixture will produce either aggregates of multiphase crystallites or uncontrolled In-rich, heavily electron-doped InSe. This fundamental thermodynamic constraint provides a compelling explanation for the large discrepancies observed in the reported physical properties of InSe. We overcome these limitations by utilizing the travelling solvent floating zone (TSFZ) method to produce high quality, centimeter-size InSe single crystals. Electrical, thermal, and thermoelectric transport measurements demonstrate that TSFZ-InSe single crystals closely**

**approach the intrinsic limit, establishing it as a benchmark material for the future studies of this important material.**

## Introduction

The realization of pure intrinsic semiconductors underlies the broad and profoundly fruitful ramifications of semiconductor science, spanning applications in electronic, photonic, energy, and quantum technologies. The availability of truly intrinsic semiconductors enables fundamental understanding of pristine material properties, which in turn forms the basis for developing core concepts such as ambipolar transport and device functionality.[1, 2] In practice, intrinsic semiconductors provide a dependable platform for controlled *n*- and *p*-type doping, which are fundamental to the design and fabrication of optoelectronic, photovoltaic, and thermoelectric devices. In parallel, the purity and crystallinity of a semiconductor are crucial in defining its quantum behaviour. The quantum Hall effect, an archetypal phenomenon that emerges only in exceptionally clean, high-mobility systems, illustrates the necessity of such material quality in accessing quantum regimes. However, achieving intrinsic and phase-pure materials remains a significant challenge, often requiring enabling advances in synthesis techniques. InSe, the van der Waals (vdW) solid being the topic of this paper, is an intriguing and promising material with non-trivial, high-performance electronic properties, further enriched by its mechanical flexibility. Unfortunately, the thermodynamics of InSe dictates that it decomposes into secondary phases before reaching the liquid state (Fig 1b), precluding straightforward crystallization from a stoichiometric melt. In this paper, we show that by using a carefully tailored synthesis, the thermodynamic difficulty can be by-passed and nearly intrinsic InSe is characterised for the first time.

InSe (Fig 1a) is a semiconductor that has motivated a long history of research for a few reasons. The most recent of them is that InSe has emerged as a flexible semiconductor, following shortly after the identification of $Ag_2S$ in 2018 as a benchmark inorganic flexible material.[3, 4] The combination of mechanical flexibility and the fast electronic response inherent to inorganic semiconductors holds promise for next-generation devices that may surpass the performance of organic counterparts, thereby revitalizing interest in InSe. The flexibility of InSe is particularly valuable in the design of wearable thermoelectric modules,[5-7] which benefit from large-area conformal contact and have potential to power pacemakers ,[8] auditory aids[9] and electrical stimulation devices.[10] One proposed origin of InSe's mechanical flexibility is that interlayer slipping is energetically more favourable than exfoliation.[4] Because the microscopic mechanisms governing this flexibility remain unclear and due to InSe's ability to retain its electronic properties under substantial deformation the study of InSe is of fundamental interest.

The discovery of InSe's flexibility is, in fact, a continuation of long-standing efforts to explore its unique properties. Early findings of an unusually high and anisotropic electron mobility[11-14] have motivated the recent observation of the quantum Hall effect in 2D-InSe[15] and advanced the understanding of its intriguing band structure.[16-20] The exceptional electron mobility is consistent with a light effective mass, arising from the steep curvature of the conduction band minimum. In contrast, the valence band maximum is highly sensitive to dimensionality and exhibits a van Hove singularity with diverging density of states (DOS) in 2D limits. InSe with *p*-type band conduction can be a compelling platform for exploring emergent low-dimensional multiferroic phenomena.[18-23]

Unfortunately, while being a rare convergence of a few important and high-potential properties, obtaining intrinsic InSe is challenging due to its peritectic thermodynamic properties. To date, efforts in synthesizing undoped InSe, both in poly- and single-crystalline morphologies, have always resulted in rather heavily *n*-doped samples showing inconsistent physical properties.[14, 24-34] This persistent inability to obtain intrinsic InSe not only hampers the investigation of non-trivial electronic states in the valence band but also limits the development of InSe-based thermoelectric modules, which require systematic control of both *n*-type and *p*-type conduction. More importantly, it leads to a vast disagreement in the values of important electronic parameters and the potential misconception that InSe is intrinsically *n*-type, as shown below.[35]

These problems originate from the fact that the widely selected synthesis route to InSe, is based on cooling a 1:1 In:Se melt, which is unstable due to the peritectic nature of InSe.[36, 37] Thermodynamically, the products of such processes can be multiphase aggregation of small crystallites or heavily In-rich samples. Uncontrollable grain-boundaries and contaminations from undesired phases resulting from fast-cooled synthesis offer reasonable explanations for the disagreement in the reported properties of InSe. In slow-cool Bridgman synthesis, the excess In occupying the interstitial sites in the structure easily donate electrons to the conduction band, and the obtained InSe is consistently *n*-type.

In this study, we show that our work overcomes the thermodynamic challenge in the synthesis, and we have obtained nearly intrinsic InSe single crystals for the first time. From the stand point of crystal growth physicochemistry,[36, 37] we tailored an optical mirror furnace

Travelling Solvent Floating Zone (TSFZ) technique,[38] which yield pure-phase and high-quality bulk InSe crystals in centimeter-size.[39] The measurements of electric and thermoelectric transport properties clearly show that TSFZ-InSe is lightly hole-doped and closely approach the intrinsic limit, establishing it as a benchmark material for the future studies of this important material.

TSFZ has been a dedicated method for growing single crystals of incongruently melted compounds such as cuprate superconductors,[40] battery,[41] and laser materials.[42] To our knowledge, our synthesis of InSe is the first study demonstrating the usefulness of TSFZ in growing high quality single crystals of a VdW material and the inclusion of volatile selenium in an optical floating zone furnace.

The structure of the paper is as follows. Immediately after the introduction, we will give a short description of the physicochemistry of the crystal growth for InSe. The thermodynamic insights provide a direct explanation for the significant inconsistencies in reported physical properties, which will be briefly reviewed in the subsequent sections. The same thermodynamic argument is the key leading to our selection of TSFZ as the synthesis method essential in producing high-quality crystals. Although we have reported the growing techniques previously,[39] the synthesis technique and the crystal quality are indispensable for understanding the observed physical properties of TSFZ-InSe. These properties, which are the central experimental results of the current paper, are described next section. The discussion section will focus on the electronic structure of our nearly intrinsic TSFZ-InSe crystal, followed by our conclusion.

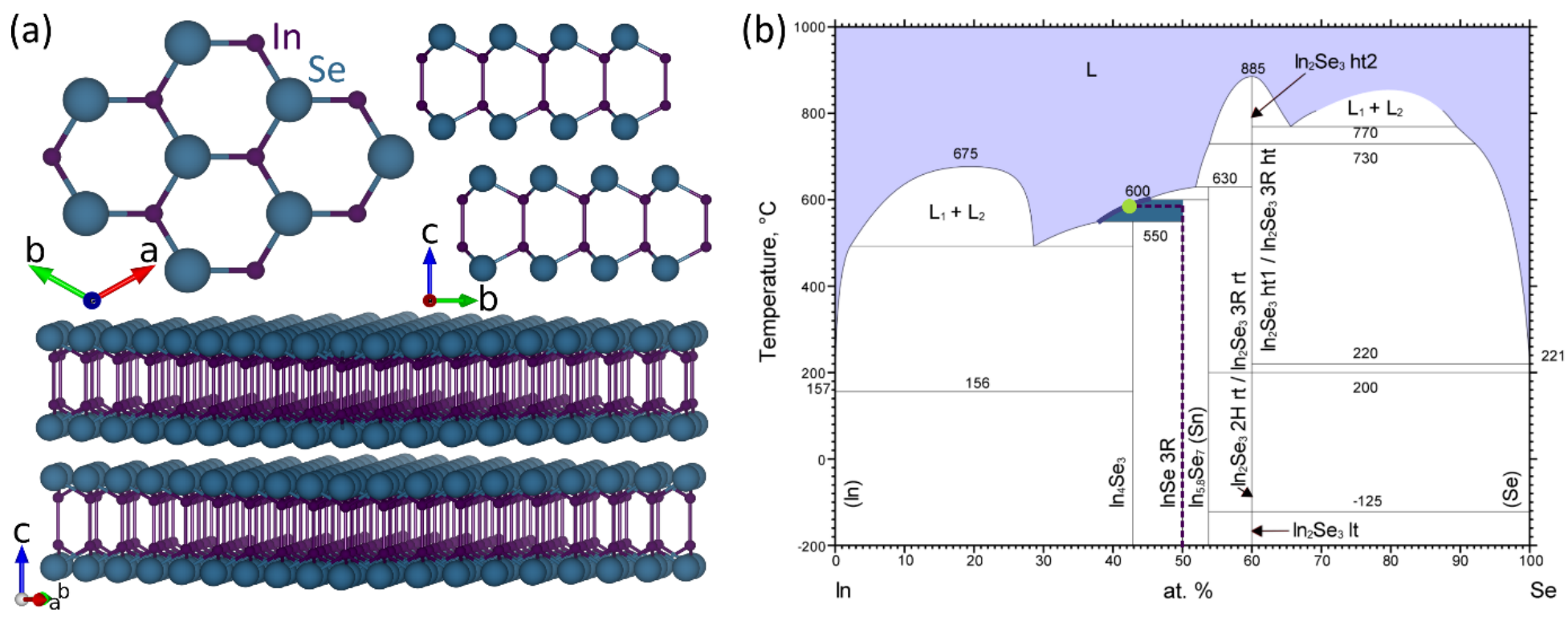

**Fig 1**. (a) Crystal structure of InSe as determined from single crystal X-ray diffraction.[39] Each layer consists of a sequence of Se-In-In-Se atoms which are separated by a vdW gap. In the ab-plane, the structure is formed of a honeycomb type lattice of In and Se. VESTA[43] was used to visualize the structure. (b) Phase diagram of InSe system adapted from Okamoto;[37] solid line shows liquidus line, circle shows the point on the liquidus line where the composition is 58:42: In:Se which allows for the growth of 1:1 In:Se. The dashed line shows the solidification path to InSe from the 58:42 In:Se composition used and shaded area shows InSe + L part of the phase diagram.

## Physicochemistry of crystal growth for InSe

Intense efforts have been made to obtain phase-pure InSe samples in both poly- and single-crystalline forms.[24-34] The commonly used synthesis route largely relies on cooling a melt of 1:1 stoichiometric In:Se. In studies aiming at thermoelectric applications, spark plasma sintering (SPS) was used.[24-31] On the other hand, the selected method for growing InSe single crystals is the Bridgman-Stockbarger method.[32-34] This method relies on the slow cooling process that also starts with a 1:1 melt. As we will show below, the widely adopted 1:1 stoichiometric precursor is thermodynamically unstable. We believe that this is the reason that, despite adopting seemingly identical synthesis routes, the observed physical properties of InSe are radically inconsistent (**Fig. 2**).

From the thermodynamic point of view, the phase diagram (**Fig. 1b**) shows that InSe melts incongruently.[37] An effort in growing InSe by slowly cooling down a 1:1 melt of In:Se would first yield $In_{5.8}Se_7$ as its most probable product. The process of making the unwanted $In_{5.8}Se_7$ would then continue until Se is depleted enough so that the remaining composition reaches 55:45 In:Se favourable for crystallizing InSe.[37] Se would then continue to deplete from the melt until a composition of 62:38 In:Se is reached and the crystallization of $In_4Se_3$ is favored.[37] In a fast cooled 1:1 melt, parts of the melt would cool rapidly and produce InSe, and the rapid cooling would result in nucleation simultaneously at multiple locations, resulting in an aggregate of small crystallites.[44] The remaining portions of the melt, which cool at various slower rates, tend to yield a variety of undesired $In_xSe_y$ phases, with compositions dependent on the local In:Se ratios.[37]

**Physical properties of InSe as reported in literature**

**Fig. 2** illustrates the reported significant discrepancies in the transport parameters of InSe, as evidenced by a comparison across different studies. In compiling **Fig.** 2, we either used the values explicitly stated in the corresponding report or extracted them from the figures with the help of PlotDigitizer.[45]

**Reported physical properties of InSe obtained by Spark Plasma Sintering**

**Fig. 2a** shows a large discrepancy in the reported values of electrical conductivity of InSe, even when comparing samples prepared using the same synthesis route. The largest conductivity reported for InSe samples prepared using a Spark Plasma Sintering (SPS) press (SPS-InSe) is about 0.2 S cm$^{-1}$,[30] while the smallest reported conductivity is about $6.7 \cdot 10^{-5}$ S cm$^{-1}$,[26] i.e. the difference is more than four orders of magnitude. The large discrepancy in electrical conductivity could be due to impurities arising from the differences in preparation methods. The sample with the smallest conductivity was prepared by mechanical alloying before SPS pressing while the sample with the largest conductivity was prepared by melting stoichiometric amounts of In and Se before SPS pressing. The reported powder X-Ray diffractogram (PXRD) for the SPS-InSe with the smallest reported conductivity suggests that at least two different phases are present in the sample.[26] These impurities could potentially cause the small conductivity observed for this sample.

The conductivities of samples produced from similar synthesis methods also vary significantly. For instance, two other samples of SPS-InSe have reported conductivities around $1.1 \cdot 10^{-4}$ S cm$^{-1}$.[24, 25] These samples were prepared through melting stoichiometric amounts of In and Se before SPS pressing. Thus, the observed conductivity of these samples, which are prepared in a similar manner to the sample exhibiting a conductivity of 0.19 S cm$^{-1}$, is found to differ by more than four orders of magnitude. From the reported data, it is also evident that the measurement direction does not influence the observed conductivity. Large discrepancies were also observed between samples which have been measured parallel to the pressing direction and perpendicular to the pressing direction. Furthermore, it is observed that the electrical conductivity measured on the SPS-InSe samples are independent of the direction of measurement as seen from the samples prepared by Lee *et al*.[27] and Shi *et al*.[25] Thus, the electrical conductivity of the SPS-InSe samples is found to be isotropic despite the anisotropic nature of InSe.

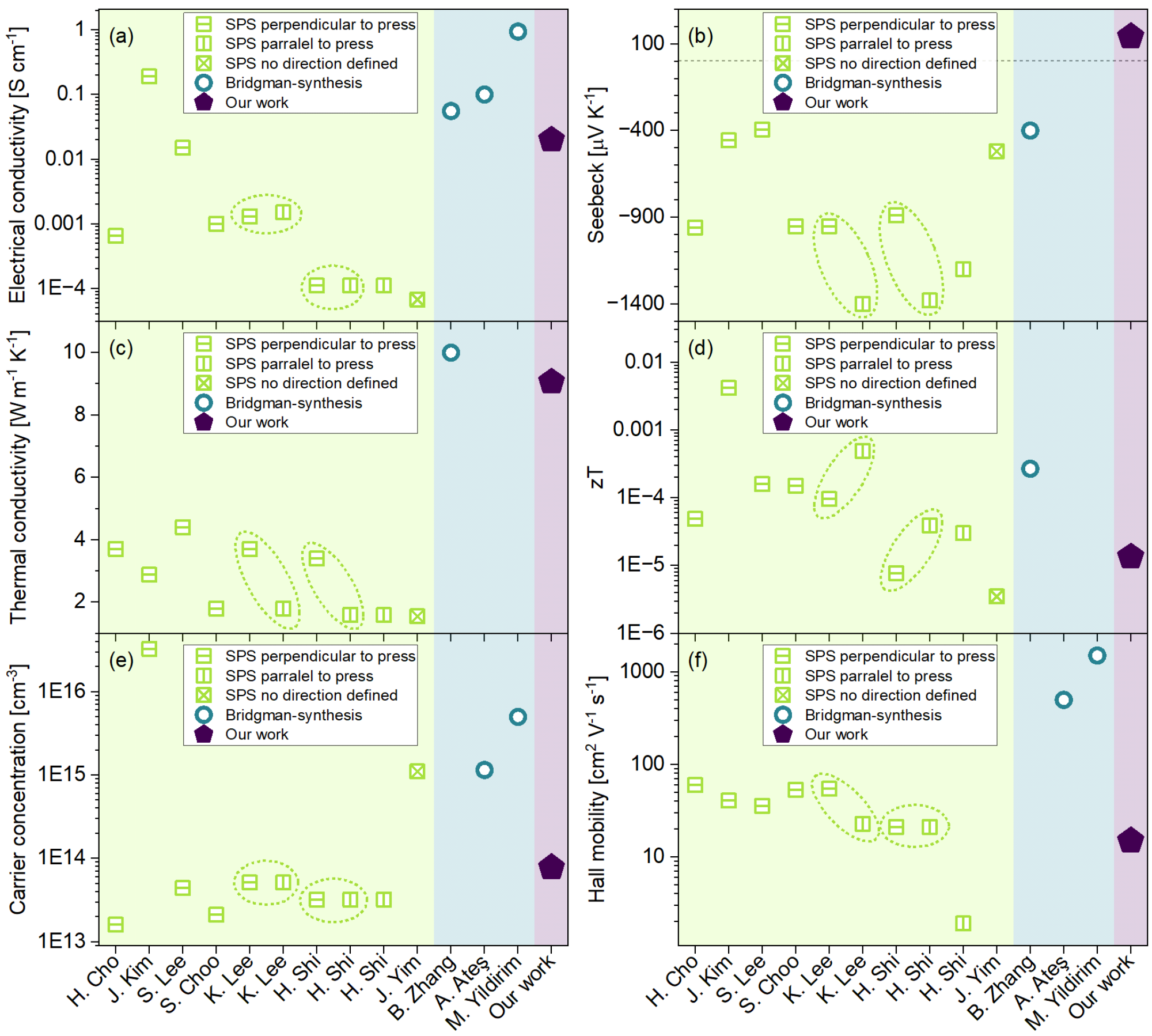

**Fig 2.** Reported room temperature ($T \approx 300$ K) values of (a) electrical conductivity (b) Seebeck coefficient, (c) thermal conductivity, (d) thermoelectric zT (e) carrier concentration, and (f) Hall mobility. Squares represent data collected on SPS-InSe samples. A horizontal line inside the square represents samples measured perpendicular to the pressing direction, a vertical line represents samples measured parallel to the pressing direction and crosses represent samples where the direction of measurement was not specified. Circles represent data measured on Bridgman-InSe samples and for all these, no mention of the measurement direction was made. Pentagons represent our data. When no symbol is present in the position corresponding to a specific reference this property was not reported in the reference. When measured values were not explicitly stated in earlier reported work the numbers were extracted from the published figures using PlotDigitizer.[45]

The large discrepancy can also be seen in the reported Seebeck coefficients (**Fig. 2b**). All values of the Seebeck coefficient are reported to be negative, indicating that the reported InSe samples are *n*-type semiconductors. For the SPS-InSe samples, the largest reported value of the Seebeck coefficient is about -1400 μV $K^{-1}$ [25] while the smallest reported value is about -394 μV $K^{-1}$.[27] Generally, the Seebeck coefficient is found to be larger when measured in the direction parallel to the pressing direction for the SPS-InSe samples suggesting some degree of anisotropy contradictory to the observations from the conductivity measurements. Lee *et al.* report PXRD measured both along and perpendicular to the pressing direction.[27] A high degree of preferred orientation is present in the pressed sample, and they therefore propose that the measurements taken parallel to the pressing direction have a greater degree of *c*-axis related properties while the measurements taken perpendicular to the pressing direction have a greater degree of *ab*-plane related properties.[27] As the electrical conductivity showed isotropic properties it is unlikely that only the proposed anisotropic nature of the sample is responsible for the difference in observed values along and perpendicular to the pressing direction.

The reported thermal conductivities (**Fig. 2c**) reflect the mutual disagreements found in electric conductivity and Seebeck coefficient. Furthermore, no conclusive display of anisotropy is observed from thermal conductivity. As can be seen, the measurements performed along the pressing direction show smaller thermal conductivity compared to the same sample measured perpendicular to the pressing direction, but these values are equal to values measured perpendicular to the pressing direction on other samples. Finally, the carrier concentration (**Fig. 2e**) and Hall mobility (**Fig. 2f**) of the SPS-InSe samples also vary strongly.

The large discrepancies observed for the SPS-InSe samples which are prepared using similar methods can likely be explained by the incongruent nature of InSe. Looking at the reported Powder X-ray Diffraction (PXRD) patterns from literature, it is evident that InSe is the majority phase in all the SPS-InSe samples.[24-31] However, some peaks which cannot be attributed to InSe are present in the SPS-InSe samples. As these peaks originate from impurity phases, they are not very intense compared to the high intensity of the (006) peak of InSe and can therefore not be expected to be assigned to specific phases when measured using in-house diffractometers. Despite being minority phases, the impurities are expected to introduce pronounced differences to the observed physical properties. Comparing two different diffractograms, it is seen that one has impurity peaks at $2\theta \approx 18°$ and $2\theta \approx 31°$,[29] while another has impurity peaks at $2\theta \approx 14°$ and $2\theta \approx 24°$[31] suggesting that the impurity phases cannot be controlled through melting stoichiometric InSe followed by SPS pressing confirming the peritectic nature of the system.

**Reported physical properties of Bridgman-InSe**

In comparison with thermoelectrics-oriented studies of SPS-InSe, InSe crystals grown by the Bridgman method have played a central role in the existing research of the quantum properties of InSe. Bridgman-InSe are currently the primary source of bulk material for producing atomically thin InSe flakes employed in opto-electronic and quantum Hall effect devices. The discovery of exotic physical properties in these few-layer systems has drawn significantly more attention than the more fundamental investigation of the parent bulk crystals, even though atomically thin flakes inevitably inherit imperfections from their bulk origins. Within few works studying bulk properties, we found that the deviation in the reported values of electrical conductivity is more than one order of magnitude. Only one study of thermoelectric properties of Bridgman-InSe was found (**Fig. 2b**), which report a Seebeck coefficient $S \approx -400$ μV · K$^{-1}$S.[32] Even without any intentional doping, Bridgman-InSe usually appears as an *n*-type semiconductor (**Fig. 2e**), with carrier number varying from $10^{15}$ cm$^{-3}$ to $10^{16}$ cm$^{-3}$.[33, 34]

As observed in SPS-InSe, such discrepancies likely originate from the peritectic nature of InSe, which in turn results in the presence of undesired impurity phases in the final crystals. Although thorough chemical and structural investigations have not been carried out for Bridgman-InSe, limited PXRD measurement exposed a shoulder appearing near the (006) peak, suggesting the existence of impurities.[32] Furthermore, energy dispersive X-ray spectroscopy measurements usually indicate that Bridgman-InSe is markedly In-rich, with In:Se ratio being typically around 51.5:48.5.[32] As suggested by density functional theory (DFT) calculations,[20, 35] the energetically most favourable defect under In-rich conditions is indium atoms occupying various interstitial sites ($In_{in}$), which give rise to shallow donor levels. In-interstitials as the main defect type corroborate the *n*-type behaviours commonly observed in Bridgman-InSe. Although *n*-type carriers in InSe are desirable due to their high mobility, this unintentional doping originates from non-ideal nucleation and crystal growth in the vicinity of the peritectic region, often resulting in uncontrolled spatial variation in composition and phase distribution in the obtained samples.

**Comparison of SPS-InSe and Bridgman-InSe**

In **Fig. 2a**, the conductivities of Bridgman-InSe samples are generally larger than those of SPS-InSe, with one exception.[30] This difference can be attributed to the larger crystallites in the Bridgman-InSe samples compared to the SPS-InSe which will result in fewer grain boundaries and therefore increase in the conductivity.[46] The thermal conductivity (**Fig. 2c**) can be explained using similar arguments. Here, the larger grains in the Bridgman-InSe sample results in larger thermal conductivity as the grain boundaries present in the SPS-InSe samples will scatter phonons travelling through the sample thereby decreasing the thermal conductivity.[47] Comparison of the carrier concentration and Hall mobility (**Fig. 2e** and **Fig. 2f**) show that the SPS-InSe samples have fewer charge carriers and a lower Hall mobility compared to the Bridgman-InSe samples.

The reported $zT$ values (**Fig. 2d**) indicate that InSe cannot be a good thermoelectric device without tuning its electrical conductivity via doping. The polarity of the intrinsic carrier type is vital in engineering InSe towards a *n*- or *p*-type semiconductor The reported Seebeck coefficients of SPS and Bridgman-Stockbager samples are always negative (**Fig. 2b**), suggesting that InSe is an *n*-type semiconductor. However, DFT calculations[4, 48] show that the Fermi level locates near the top of the valence band, suggesting an intrinsic *p*-type behaviour. The disagreement between the experimental observations, made on SPS and Bridgman-Stockbager samples, and theoretical calculations raise an important question on the true intrinsic behaviours of InSe.

**Growing InSe crystals from Travelling Solvent Floating Zone (TSFZ)**

The binary phase diagram in **Fig. 1b** shows that the region where InSe solid coexists with the liquid phase, and thus allows crystal growth, occupies a rather small area in the phase diagram. The circle marks the 58:42 In:Se composition from where InSe can be grown while the dashed line shows the path from the liquidus line at the composition of In:Se 58:42 to the 1:1 In:Se composition. At this point, if no parameters of the growth are changed, InSe will crystallize and the Se content in the liquid will diminish. If the temperature is lowered, crystallization will continue to occur along the liquidus line (blue) until the composition of the melt reaches a point where crystallization of $In_4Se_3$ will occur instead.

Therefore, an effective synthesis route for InSe should necessarily start from a melt with composition being different from the 1:1 stoichiometry. In our experiment, an In:Se composition of 58:42 was selected. In principle, a different composition within the range marked by the liquidus line in **Fig. 1b** can also yield the desired InSe. However, the synthesis process is very time consuming, and only one composition was chosen in our synthesis.

The peritectic nature of InSe calls for a synthesis method that allows a continuous addition of fresh materials to ensure that the correct stoichiometry is kept throughout growth process. The TSFZ method utilizes a so-called solvent zone which has the composition (58:42 In:Se) necessary to crystallize the 1:1 In:Se phase. Fresh materials are continuously fed into the zone while crystallization occurs. The material fed to the zone has the same composition (1:1 In:Se) as that of the desired phase material, thereby a constant composition of the zone is ensured.[39]

In **Fig. 3** the as grown InSe crystal is shown; here the different parts of the crystal growth process are marked. The polycrystalline powder used as a starting material can be seen. Above the polycrystalline powder the part of the sample where grain growth happens can be seen at the bottom of the ingot. The different polycrystalline seeds grow larger until the third area is reached. The third area shows the single crystal domain of approximately 3 cm in length and 1 cm diameter. The fourth area shows the solvent zone which is solidified because of rapid cooling; the frozen solvent zone is not a single crystalline grain which is expected due to the rapid cooling.

We have confirmed the crystallinity and structure of TSFZ-InSe using the combination of PXRD, SCXRD and SEM-EDX.[39] In short, the results indicate that TSFZ-InSe crystals are phase-pure and have high crystallinity. As determined by single crystal diffraction, the space group of TSFZ-InSe is $R3m$, being consistent with the structure of the bulk γ-InSe with the lattice constants $a = 4.0069$ Å and $c = 24.967$ Å. The symmetry of each InSe quadruple layer is described by the point group $D_{3h}$, corresponding to the 2D α-polytype.[21] We also observed signatures of stacking faults, as expected for van der Waals materials. More importantly, SEM-EDX analyses show an almost stoichiometric composition with an In:Se ratio of 50.3:49.7, uniformly distributed across the single crystal region shown in **Fig. 3**.

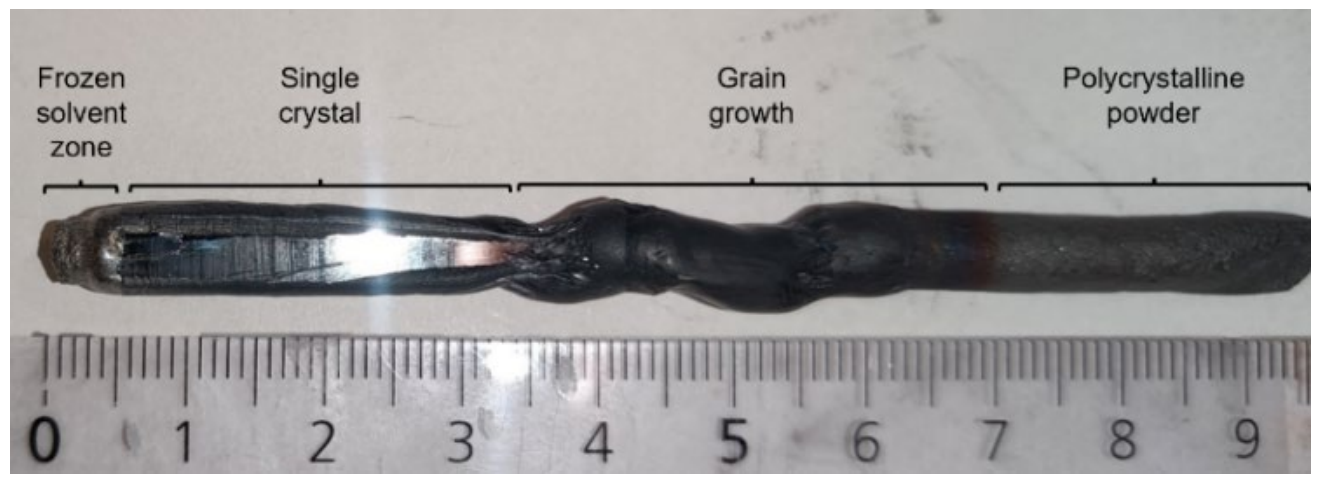

**Fig. 3**. As-grown InSe crystal after removal from the optical mirror furnace.

**Physical properties of TSFZ-InSe single crystals**

**Fig. 4a-d** summarize the in-plane transport properties the TSFZ-grown InSe crystals in the temperature window of $250\ \mathrm{K} \leq T \leq 400\ \mathrm{K}$; measurements of the out of plane properties of single crystalline InSe have proved impossible due to the soft nature of the single crystals. In general, InSe exhibits properties expected from a conventional *p*-type semiconductor, in sharp contrast to the reported *n*-type behaviours discussed earlier. As shown in **Fig. 4a**, the conductivity of InSe (σ) is about $0.02\ \mathrm{S \cdot cm^{-1}}$ at room temperature, and the conductivity is found to decrease exponentially with $1/T$. **Fig. 4b** shows that the Peltier coefficient $\Pi \equiv ST$, where $S$ is the Seebeck coefficient, is positive in the whole temperature range, suggesting that InSe is a *p*-type semiconductor. We also find that TSFZ-InSe is a good thermal conductor; its thermal conductivity (κ) at RT is about $9\ \mathrm{W \cdot (K{\cdot}m)^{-1}}$, being comparable to that of bismuth.[49] The clear crystalline peak at low temperatures conforms with the high quality of the crystal. The Seebeck coefficient is considerable, $S \approx 1.44\ \mathrm{mV \cdot K^{-1}}$ at RT, because of its large κ and small σ, meaning that InSe is a poor thermoelectric material with a Figure-of-merit $zT \approx 1.36 \times 10^{-6}$ at RT (**Fig. 4d**).

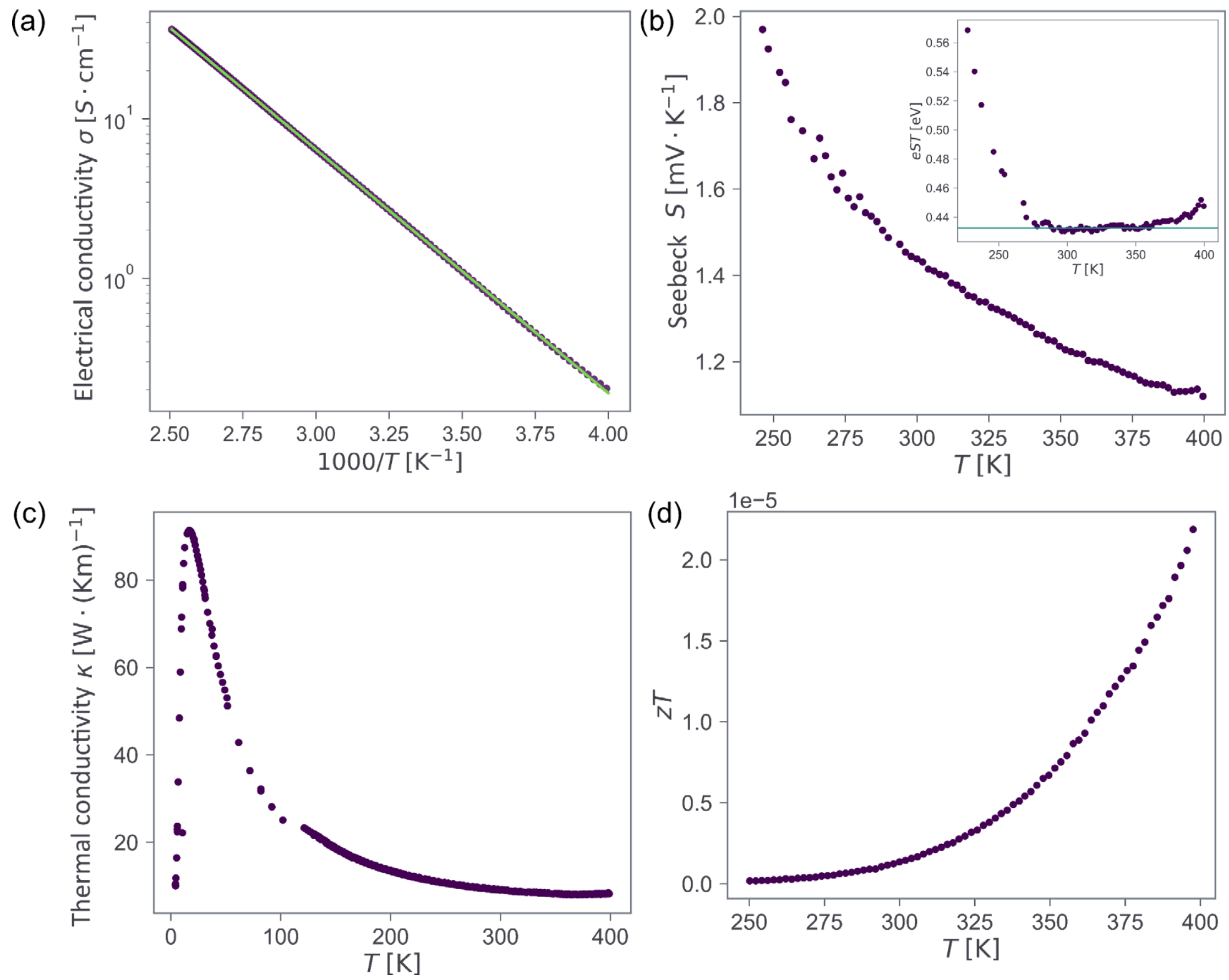

**Fig. 4.** In-plane transport properties of TSFZ-InSe single crystal: (a) Electrical conductivity as a function of temperature. The linear line shows the fit to Arrhenius law. (b) Seebeck coefficient as a function of temperature. The inset shows the temperature dependence of Peltier coefficient $\Pi = ST$. The horizontal line in the inset shows the estimation of the position of the chemical potential μ (see text). (c) Thermal conductivity. (d) Figure-of-merit $zT$.

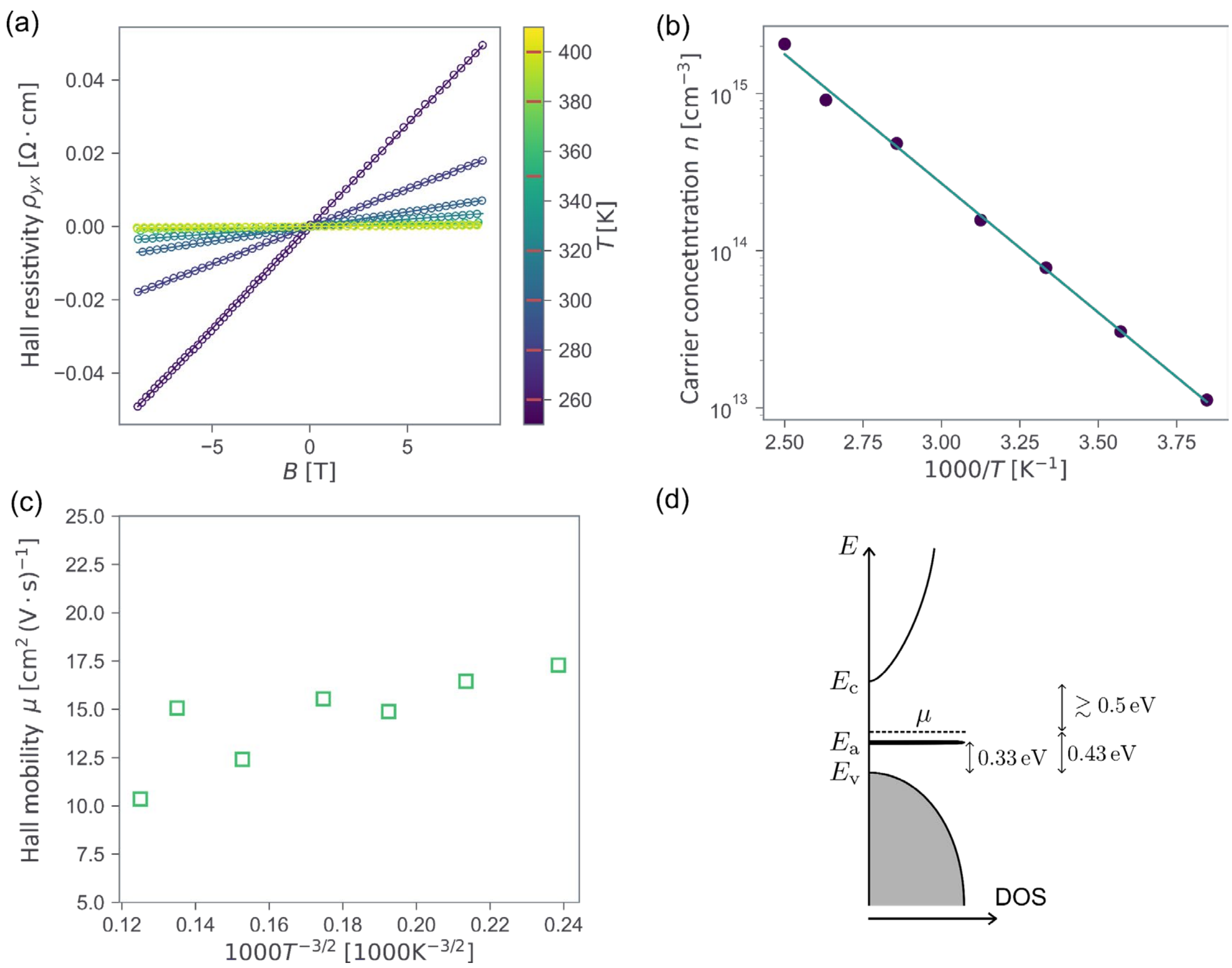

**Fig. 5.** (a) Hall resistivity of TSFZ-InSe as a function of magnetic field $B$ measured at various temperatures. The circles depict data, and the lines are linear fits. (b) The concentration of hole-like carrier as a function of $1/T$. The linear line shows the Arrhenius fit for estimating the acceptor activation energy $\Delta = E_\mathrm{a} - E_\mathrm{v}$ (see text). and mobility obtained from Hall analyses. (c) Electronic structure of TSFZ-InSe derived from transport properties (see text).

**Fig. 5a** shows that the Hall resistivity ($\rho_{yx}$) of TSFZ-InSe is positive and linear with respect to magnetic field ($B$) in the range $-9\ \mathrm{T} \leq B \leq 9\ \mathrm{T}$ at all measured temperatures. A combination of linear fits using $\rho_{yx} = 1/R_\mathrm{H}$ and the definition of Hall coeficient $R_\mathrm{H} = 1/(ne)$ provides an estimation of the hole number $n$. The hole mobility ($\mu$) can be estimated using the Drude formula $\sigma = en\mu$; $n$ and $\mu$ are shown in **Fig. 5b** as a function of temperature. The carrier number decreases exponentially with $1/T$, whereas the mobility does not show a clear $T$-dependence. At RT, $n \approx 7.8 \times 10^{13}$ cm$^{-3}$ and $\mu \approx 15$ cm$^2 \cdot (\mathrm{V} \cdot \mathrm{s})^{-1}$. The mobility weakly increases with decreasing $T$, agreeing with the $T^{-3/2}$ behaviour of semiconductors in the clean limit, where phonon scattering processes dominate the electronic transport.

**Discussion**

The positive Seebeck coefficient and the positive linear Hall effect clearly indicate that hole-like carriers are the major carrier type in single crystalline TSFZ-InSe. More importantly, the small carrier number is of the same order as those observed in very lightly hole-doped semiconductors. In fact, the experimentally estimated Hall mobility of hole-like carriers, $\mu \approx 15$ cm$^2(\mathrm{V} \cdot \mathrm{s})^{-1}$, is consistent with the theoretical value for valence band conduction in pristine InSe,[17] being a few orders of magnitude smaller than the mobility of electrons due to the much heavier effective mass. Due to the almost 1:1 stoichiometric composition, TSFZ-InSe is nearly free from $\mathrm{In}_{in}$ shallow donors. Consistently, the hole-like carrier number, $n \approx 7.8 \times 10^{13}$ cm$^{-3}$, is orders of magnitude smaller than the electronlike carrier number typically observed in Bridgman-InSe.[33, 34] Being free from unintentional electron doping, the electronic properties of TSFZ-InSe are thus likely governed by *p*-type band conduction, which so far has never been observed in InSe.

The width of the semiconducting band gap $E_g$ of γ-InSe has been the topic of a few theoretical calculations, which produced values varying from $0.8$ eV and $1.4$ eV.[20-22] In the temperature window $250$ K $\leq T \leq 400$ K, $E_g$ exceeds thermal energy $k_{\mathrm{B}}T$ by two orders of magnitude. The large band gap and the value of $n$ suggest that the electronic conduction of TSFZ-InSe is not in the intrinsic regime at RT. More specifically, at $T = 300$ K, by using $E_g = 1$ eV and putting the intrinsic chemical potential $\mu_i$ at 0.5 eV above the edge of the valence band $E_\mathrm{v}$, the formula $n = N_\mathrm{v} \exp(-\beta(\mu_i - E_\mathrm{v}))$,[1] with $\beta \equiv 1/(k_\mathrm{B}T)$, yields $N_\mathrm{v} \approx 9.3 \times 10^{23}$ cm$^{-3}$ as the effective density of states at the valence band maximum. This value is clearly over-exaggerated, especially when comparing with a more realistic estimation using the formula $N_\mathrm{v} = 2.5(m_\mathrm{h}/m_0)^{3/2}(T/300\ \mathrm{K})^{3/2} \times 10^{19}$ cm$^{-3}$,[1] and the calculated value of effective mass of holes $m_\mathrm{h} = 2.6m_0$,[17] which produces a reasonable $N_\mathrm{v} \approx 4.03 \times 10^{19}$ cm$^{-3}$. The small number of hole-like carriers in TSFZ-InSe therefore likely comes from an acceptor-like impurity. Since no evidence of foreign impurities was detected in our compositional analysis,[39] the observed acceptor level is likely associated with a native defect, as is commonly the case in compound semiconductors.

The exponential dependency on $1/T$ found in both $\sigma$ and $n$ in **Fig. 4a** and **Fig. 5b** indicate that around RT, the ionization of both acceptor and donor states in TSFZ-InSe occur only partially, i.e. the material is in its low temperature regime.[1] The conduction is then dominated by holes thermally excited from the acceptor level $E_\mathrm{a}$ to the edge of valence band $E_\mathrm{v}$, and thus the energetic difference $\Delta = E_\mathrm{a} - E_\mathrm{v}$ defines the temperature dependence of $n$.[35] Fitting $n(T)$ to the rule $n(T) = \frac{N_\mathrm{v}(N_\mathrm{a}-N_\mathrm{d})}{N_\mathrm{d}} \exp(-\beta\Delta)$ yields $\Delta \approx 0.33$ eV and $\frac{N_\mathrm{v}(N_\mathrm{a}-N_\mathrm{d})}{N_\mathrm{d}} \approx 2.29 \times 10^{19}$ cm$^{-3}$. Using $N_\mathrm{v} \approx 4.03 \times 10^{19}$ cm$^{-3}$ as estimated above, we obtain $N_\mathrm{a}/N_\mathrm{d} \approx 1.22$ as the ratio between the densities of the acceptor and donor impurities. On the other hand, the Peltier coefficient $\Pi$ can be interpreted in terms of the distance between the chemical potential $\mu$ and $E_\mathrm{v}$, i.e., $e\Pi \approx \beta(\mu - E_\mathrm{v})$.[50, 51] Using the constant region in the $\Pi(T)$ curve, we estimate that $\mu$ is about 0.43 eV above $E_\mathrm{v}$, i.e., at the vicinity of the intrinsic mid-gap position. The combined information from Hall and Peltier effects allows us to derive the electronic structure illustrated in **Fig. 5c**. We note that the distance between the conduction band minimum $E_c$ and μ is approximated using $E_g \approx 1\,\text{eV}$ as suggested from DFT calculations.[15] TSFZ-InSe is an almost intrinsic semiconductor with the presence of a small number of deep, acceptor-like defects. The near-unity ratio of $N_\text{a}/N_\text{d}$ further implies that a single species of native defect plays the dual role as both acceptor and donor.

Among the various native defects investigated by DFT calculations,[20, 35] $In_{in}$ has the lowest formation energy of about $1\,\text{eV}$ and is the shallow donor that often dominates the electronic properties of InSe. Chemical composition analyses and Hall effect clearly show that this defect is unimportant in TSFZ-InSe. Selenium vacancy ($V_{Se}$) is almost as easy to form; its formation energy is about $50\,\text{eV}$ larger than that of $In_{in}$.[20, 35] Interestingly, $V_{Se}$ is amphoteric in nature and has high ionization energies, creating deep acceptor and donor levels.[20] We thus attribute $V_{Se}$ as the dominant species of defects in TSFZ-InSe. Within the framework of our analyses, we cannot estimate the defect concentration accurately. However, the small hole number $n$ strongly suggests a very low concentration of these defects, being at the same order of magnitude as $n$, i.e. about $10^{13}$ cm$^{-3}$.

As mentioned above, melting growth techniques such as the Bridgman method are thermodynamically unstable and always yield In-rich, *n*-type InSe with a rather large number of electronlike carriers ($n{\sim}10^{15}$ cm$^{-3}$) likely originating from $In_{in}$ defects. TSFZ-grown, nearly intrinsic InSe overcomes this longstanding quality bottleneck, thereby paving the way for expansion and deepening of fundamental knowledge on this important semiconductor. For instance, using devices fabricated from flakes exfoliated from In-rich Bridgman-grown crystals, the quantum Hall state[15] has been observed as a hallmark of the high mobility of electrons in InSe.[12, 17] Equally intriguing to the high mobility electron band is the Mexican-hat shape of the valence band edge in the 2D limit of InSe.[19, 20] The divergent DOS arising from this distinctive band structure can promote a variety of two-dimensional multiferroic orderings tuneable via electrostatic gating.[19] However, approaching the valence band conduction by gating has been challenging because of the large number of *n*-type carriers inherent to In-rich samples. Similarly, a process for controlled *p*-type doping of InSe,[35] being vital for making flexible electronic and thermoelectric devices, has proved difficult to achieve because of carrier compensation effects. The nearly intrinsic TSFZ-InSe crystals can provide a reliable platform for both studies.

## Conclusion

InSe is a vdW solid that can be easily exfoliated to approach the 2D limit and it has excellent mechanical flexibility. Its electronic properties benefit from both high-mobility electrons in the conduction band and a non-trivial structure in the valence band. Because of its combined merit, InSe is an important semiconductor holding a variety of intriguing properties and promising applications. The stringent thermodynamic requirements for crystal formation have long hindered the synthesis of phase-pure InSe, thereby limiting its potential. The reported parameters of the materials are highly inconsistent due to the difficulty in establishing a reliable synthesis route. Guided by the physicochemistry of the In:Se binary system, we show that the travelling solvent floating zone method is a controllable and effective route to grow high quality γ-InSe single crystals. The measurements of electronic properties show that InSe grown using the travelling solvent floating zone method most closely approach the intrinsic quality expected of ideal InSe. These results establish InSe grown using the travelling solvent floating zone method as a benchmark material for uncovering the intrinsic physics of InSe and for enabling its full potential in future electronic and optoelectronic applications.

**Acknowledgements**

The work was supported by the Villum Foundation (25861), the Danish National Research Foundation (DNRF189), the Carlsberg Foundation (CF24-1527), the Novo Nordisk Foundation (NNF23OC0085064) and the Independent Research Fund Denmark.

**Conflict of interest**

The authors declare no conflict of interest.

## Supporting Information

# Intrinsic physical properties of flexible van der Waals semiconductor InSe

Jacob Svane,[a] Khuong Kim Huynh,[a*] Yong P. Chen,[b] and Bo Brummerstedt Iversen[a*]

[a]Center for Sustainable Energy Materials, Department of Chemistry, Aarhus University, 8000 Aarhus C, Denmark

[b]Department of Physics and Astronomy, Aarhus University, 8000 Aarhus C, Denmark

E-mail: hkkhuong@chem.au.dk, bo@chem.au.dk

### Experimental measurements

### Electrical measurements

The resistivity, Hall effect, and Seebeck coefficient measurements were carried out on a $1.8 \times 0.87 \times 0.15$ mm$^3$ single crystal cleaved from single crystalline region of the as-grown ingot. Here the shortest dimension of the sample is parallel to the *c*-axis, and the electric or heat current was applied parallel to the *ab*-plane. For the measurements of the electrical transport properties, six $\phi 25$ μm gold wires were glued to the sample via conducting carbon paste. The measurements were carried out using an Electrical Transport Option (ETO) for Quantum Design Physical Properties Measurement System (PPMS). Additional measure-ments using equivalent lock-in technique and Delta DC method via home-built systems were also carried out. All measurements yield identical temperature and magnetic field dependencies of resistivity and Hall effects.

### Seebeck coefficient

In the thermoelectric measurements, two E-type thermocouples were additionally connected to the sample, and the whole sample setup was bridged across two heaters. The Delta-averaged Seebeck voltage is calculated as $V_S = 0.5 \times [V_S(+) - V_S(-)]$. Here $V_S(+)$ is the Seebeck voltage measured with the first heater turning on and the second heater turning off, and $V_S(-)$ is the Seebeck voltage arising from the heat current flowing in the reversed direction. The purpose of this setup is to eliminate the parasitic thermoelectric effect that may be present in the measurement circuit.

### Thermal conductivity

For thermal conductivity measurement, a large chunk of the as-grown ingot was cast in resin and then cut by a precision diamond saw at minimum cutting speed and pressure. A $8$ mm-long sample with the cross-section of $1.31\,\mathrm{mm}^2$ was carefully mounted to the puck for PPMS's Thermal Transport Option (TTO). The heat current was applied along a direction almost parallel to the *ab*-plane.